\documentclass[pra,twocolumn,showpacs,preprintnumbers,amsmath,amssymb,floatfix,superscriptaddress] {revtex4-1}

\usepackage{graphicx}
\usepackage{dcolumn}
\usepackage{bm,bbm}
\usepackage{color}
\usepackage{textcomp}

\begin{document}


\title{Wavelength dependence of laser pulse filamentation in the close spectral vicinity of atomic resonances }

\email{demeter.gabor@wigner.hun-ren.hu}
\author{G. Demeter}

\affiliation{HUN-REN Wigner Research Centre for Physics, Budapest, Hungary}





\date{\today}

\begin{abstract}
We investigate the propagation and nonlinear self-focusing of TW power laser pulses that create 10-m-scale, highly homogeneous plasma channels in rubidium vapor.  Using computational solutions of the relevant propagation equations, we study the effects of the ionizing pulse central wavelength in relation to the resonance frequencies of atomic rubidium. Recent experiments show that pulse propagation and plasma channel creation is distinctly different for 780 nm laser pulses (resonant with the rubidium $\mathrm{D}_2$ line) and 810 nm laser pulses. We study pulse propagation in a $\pm$30 nm range around the $\mathrm{D}_2$ resonance and find that the results are distinctly different when tuning to sub-resonant wavelengths from those obtained for super-resonant wavelengths. For pulse wavelengths below the resonance the behavior is similar to the resonant case, characterized by strong self-focusing and a sharp plasma boundary. Pulse wavelengths above 780 nm on the other hand yield gradually weaker self-focusing and an increasingly diffuse plasma boundary. Our results suggest that the observed behavior can be attributed to an interplay between  multiple factors: anomalous dispersion around atomic resonances, resonant transitions between excited states of Rb lying in the 740-780 nm range and wavelength-dependent multiphoton ionization rates.
\end{abstract}


\maketitle

\section{Introduction}
\label{intro}

The propagation of high-power laser pulses in gaseous media gives rise to a variety of nonlinear optical phenomena, including self-focusing, filamentation, plasma channel formation, self-phase modulation, and supercontinuum generation \cite{Couairon2007, Berge2007, Kandidov2009}. These effects have found widespread application in fields such as ultrafast spectroscopy, high-harmonic generation, remote sensing, and laser-induced breakdown spectroscopy \cite{Corkum2007, Kasparian2008}. One specific, but very important application of plasma channels formed by high-power laser pulses is in 
wakefield acceleration, a next-generation particle acceleration technique that offers the potential for significantly more compact accelerators. In Laser Wakefield Acceleration (LWFA), large-amplitude plasma waves—referred to as wakefields—are excited by the extremely large electric fields of laser pulses, while in the Plasma Wakefield Acceleration (PWFA) scheme,  they are driven by a relativistic charged particle beam propagating through a plasma. These wakefields can support accelerating gradients several orders of magnitude greater than those achievable in traditional RF accelerators \cite{Chen1985, Blumenfeld2007, Joshi2003, Esarey2009, Leemans2009, Tajima2020}. This technique requires a pre-formed plasma medium, typically created by ionizing a neutral gas using terawatt (TW) or petawatt (PW) class laser pulses.

One notable PWFA project is the Advanced Wakefield Experiment (AWAKE) at CERN, a wakefield accelerator driven by high-energy proton bunches \cite{Caldwell2009,Gschwendtner2016,Adli2018v2}. AWAKE is unique in many respects, one of which is the requirement of a long, 10-meter-scale plasma channel with precisely engineered density distribution for the wakefield generation to function properly over the entire length of the accelerator structure \cite{Lotov2013}. As such, it employs a very specific scenario for the generation of the plasma: rubidium vapor with precisely controlled density distribution is prepared first, then subjected to a $\sim 120$ fs, TW power laser pulse that removes the single valence electron of Rb with a probability very close to 1, leaving the inner, closed electron shells intact \cite{Oz2014, awake2013design}. Ionization is facilitated by the fact that the ionizing pulse propagating in Rb vapor has a 780 nm central wavelength, resonant with the $\mathrm{D}_2$ transition of atomic Rb from the ground state. 

This propagation scenario with a strong single-photon resonance is fairly special in that resonant transitions provide a powerful nonlinear feedback on the laser field. The phenomena this feedback gives rise to has been studied initially for much lower field intensities in resonant nonlinear optics \cite{Lamb1971, Lamare1994, Delagnes2008}. The feedback due to resonant transitions, however, ceases once the valence electron has been removed from the atom. As the vapor density in the AWAKE rubidium vapor source ($\mathcal{N}=1-10 \times10^{14}\mathrm{~cm^{-3}}$) is much smaller than the atmospheric density \cite{Plyushchev2017}, the TW power laser is far below the threshold for regular self-focusing to occur, which means that once the valence electron has been removed from all the atoms, the medium is essentially transparent.
This is quite different from the usual scenarios of laser filamentation and has been studied only relatively lately \cite{Demeter2019, Demeter2021}. Recently, experiments have been performed to study the propagation of 780 nm {\em resonant} laser pulses and that of 810 nm {\em off-resonant} ones in a 10-meter-long rubidium vapor under identical conditions \cite{Demeter2024}. Measurements of the transmitted laser pulse properties and the observation of the plasma channel properties showed that there is indeed a considerable difference between pulses similar in all respects except for the central wavelength. Resonant pulses create a plasma channel with a much sharper boundary and suffer a lower energy loss for given plasma channel length. This suggests that they are focused more effectively by the rubidium vapor than off-resonant pulses.

Further study of the subject is important for several reasons. First, during the Run 2 phase of AWAKE a new setup with two consecutive 10-meter-long plasma sections are planned, with further plans to scale up the second plasma to greater lengths \cite{Gschwendtner2022}. For evaluating the technical possibilities and aiding the design of future accelerator devices potentially using plasma lengths of 100s of meters, it is vital to understand the extended pulse propagation scenarios that are envisioned. 
Second, laser pulse filamentation in the vicinity of an atomic resonance is an interesting fundamental question in its own right. It may have applications in the creation of long plasma channels for lightning protection \cite{Houard2023} or remote sensing applications in the atmosphere where resonances play an important role.

In this paper, we study the wavelength dependence of laser pulse propagation in rubidium vapor around the 780 nm wavelength $\mathrm{5^2S_{1/2}}\rightarrow\mathrm{5^2P_{3/2}}$ transition of the valence electron. We devise a model description of the rubidium atom that takes into account the ground state and, in addition, 9 important excited state levels. This model can be used to treat resonant and off-resonant interaction with the valence electron in a unified way and calculate the transient optical response in the vicinity of the $\mathrm{D}_2$ line.
We use numerical simulations of the propagation equations to calculate the evolution of laser pulses in gaseous rubidium, as well as the resulting plasma channel profile along the vapor. We first present propagation calculations as a function of input pulse energy for three wavelengths, the 780 nm resonant wavelength and the 750 nm and 810 nm wavelengths detuned by 30 nm-s from the $\mathrm{D}_2$ resonance. Next, we present calculations for fixed input energy and a varying wavelength around 780 nm. In interpreting the results, we discuss the role of an anomalous dispersion curve around atomic resonances from the ground state, the role of the wavelength dependence of multiphoton ionization rates and the effects of transitions between excited states of the valence electron
with wavelengths in the 740 nm - 780 nm range.

\section{Equations and numerical methods}

Our objective is to derive propagation equations appropriate for analyzing the long-range ($\sim$10 m) propagation of TW power, 120 fs duration laser pulses in rubidium (Rb) vapor. The pulse intensities considered are sufficient to induce multiphoton or tunnel ionization of the single valence electron from the ground state, as well as from excited electronic states. No ionization is possible from the closed 4P shell at the intensities considered, as the ionization potential for the second electron (27.29 eV) is much greater than that for the valence electron (4.18 eV). 
The vapor density ($\sim10^{14}$–$10^{15}~\mathrm{cm^{-3}}$) is orders of magnitude lower than that of air at atmospheric densities, so traditional self-focusing or filamentation effects\textendash common at similar power levels in air or noble gases\textendash are not expected to play a significant role. The central wavelengths of the pulses are assumed to lie within the 750–810 nm range, which encompasses several atomic transition lines of the valence electron. As a result, it is essential to develop a model that treats both resonant and near-resonant excitation within a unified framework.

\subsection{Field propagation equation}

The basic equation used for the pulse propagating along the $z$ axis is formulated for the complex envelope $\mathcal{E}(r,z,t)$ of an axisymmetric laser field:
\begin{equation}
 E(r,z,t)=\frac{1}{2}\mathcal{E}(r,z,t)\exp(ik_0z-\omega_0t)+c.c.
\end{equation}
where $\omega_0$ and $k_0$ are the central angular frequency and wave number in vacuum, and $r$ denotes the radial coordinate. Medium response is similarly expressed in terms of complex envelope functions: $\mathcal{P}(r,z,t)$ (atomic polarization), $\mathcal{R}(r,z,t)$ (plasma dispersion), and $\mathcal{Q}(r,z,t)$ (ionization-induced loss), all defined below.

Transforming to a reference frame co-moving with the pulse, defined by $\xi = z$ and $\tau = t - z/c$, and applying the paraxial approximation, we express the wave equation in frequency space:
$\tilde{\mathcal{E}}(\vec{r},\xi,\omega)=\mathfrak{F}\{\mathcal{E}(\vec{r},\xi,\tau)\}$, etc.
where $\mathfrak{F}\{\cdot\}$ denotes the time-Fourier transform. To accurately model ultrashort pulses and the possible formation of sharp leading edges, we adopt the Slowly Evolving Wave Approximation (SEWA) \cite{Brabec1997,Couairon2011}. The resulting propagation equation reads:
\begin{equation}
\begin{aligned}
\partial_\xi \tilde{\mathcal{E}}  = & 
\frac{i}{2k} \nabla_\perp^2\tilde{\mathcal{E}}
+ i\frac{k}{2\epsilon_0} \tilde{\mathcal{P}} \\
& -\eta_0\hbar\omega_0\mathcal{N}\tilde{\mathcal{Q}} 
- \frac{ik}{2}\frac{e^2\mathcal{N}}{\epsilon_0 m_e (\omega_0+\omega)^2} \tilde{\mathcal{R}}
\end{aligned}
\label{waveeq}
\end{equation}
where $k = (\omega_0 + \omega)/c$, $e$ and $m_e$ are the elementary charge and electron mass, $\epsilon_0$ is the vacuum permittivity, $\eta_0$ is the vacuum impedance, and $\mathcal{N}$ is the atomic density (vapor density). 

\begin{figure}[htb]
\includegraphics[width=\columnwidth]{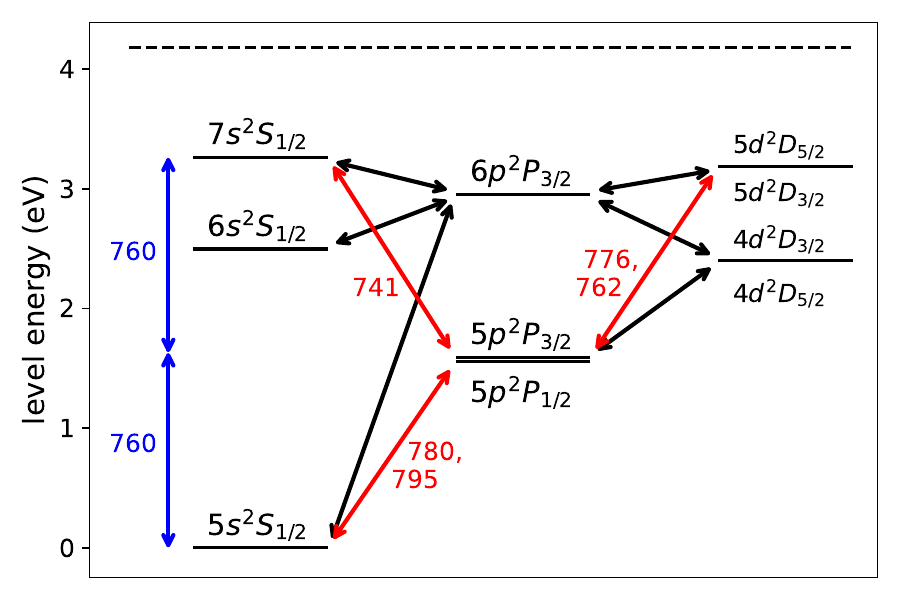}
\caption{a) Electronic levels of the rubidium atom that are included in the theoretical model. { Horizontal lines labeled above mark atomic levels, lines labeled above and below mark groups of fine-structure sublevels that are not resolved on the scale of the figure. Red and black arrows mark allowed dipole transitions between the states, or groups of transitions (transitions between different fine-structure sublevels are not resolved on the figure). Red arrows labeled with wavelength values mark transitions or groups of transitions that come close to resonance with $\lambda=750-810\mathrm{~nm}$ light. Blue lines mark a 760 nm two-photon transition. Energy is measured relative to the ground state energy, horizontal dashed line marks the ionization limit.}}
\label{fig_Rblevels}       
\end{figure}

The first term on the right-hand side (RHS) of Eq. \ref{waveeq} represents diffraction. The physical interpretation of the other three terms are as follows: The second term on the RHS is due to  the atomic polarization $\mathcal{P}$, which, for wavelengths in the immediate vicinity of an atomic resonance, will be dominated by Rabi-oscillation type transitions.  Contributions
of this type cannot be expressed in terms of usual optical nonlinear coefficients \cite{BoydNonlinOptics}. To account for this, we explicitly compute the atomic state evolution by solving the time-dependent Schr{\"o}dinger equation for the probability amplitudes $\alpha_j(t)$ in the basis of energy eigenstates:
\begin{equation} 
 |\psi\rangle=\sum_j\alpha_j(t)\exp(-i\omega_jt)|j\rangle
\end{equation}
where $\hbar\omega_j$ is the energy of $|j\rangle$. The evolution equations, written using the rotating wave approximation are given by:
\begin{equation}
\begin{aligned}
 \partial_t\alpha_j = & \frac{i}{2\hbar}\sum_k \mathcal{E}e^{-i\Delta_{jk}t} d_{jk}\alpha_k \\
 & + \frac{i}{2\hbar}\sum_{k'} \mathcal{E}^*e^{i\Delta_{jk'}t} d_{jk'}\alpha_{k'}-\frac{\Gamma_j}{2}\alpha_j
 \end{aligned}
 \label{schrodinger}
\end{equation}
where 
\begin{itemize}
 \item $d_{jk}$ are electric dipole matrix elements.
 \item The index $k$ runs over states with $\omega_k < \omega_j$ for which $d_{jk} \neq 0$. These terms represent photon absorption.
 \item The index $k'$ runs over states with $\omega_k' > \omega_j$ for which $d_{jk} \neq 0$. These term represent photon emission.
 \item $\Delta_{jk} = \omega_0 - (\omega_j - \omega_k)$ is the detuning of the pulse central frequency from the $|k\rangle \rightarrow |j\rangle$ transition.
 \item $\Gamma_j$ are ionization-induced loss rates. The (intensity dependent) multiphoton ionization rates from the ground and first excited states are evaluated using the PPT (Perelomov–Popov–Terent’ev) theory \cite{Perelomov1966, Perelomov1967, Perelomov1967b}. The explicit form used is taken from \cite{Bejot2008}, and are reproduced in the appendix of \cite{Demeter2019} for reference. Single-photon ionization rates are obtained from experimental data \cite{Duncan2001}. 
\end{itemize}
The atomic level structure and the transitions used in in extended numerical scans are depicted in Fig. \ref{fig_Rblevels}. This set of levels was chosen by executing trial runs using a much wider set of levels and then discarding those which were found to contribute only marginally to the final result.
Required atomic data\textendash energy levels and dipole matrix elements\textendash are taken from standard sources \cite{SteckRb85,Safronova2004,NIST}. The atomic transitions that may be close to resonance with the laser in our calculations are the following (wavelengths rounded to nm): 
\begin{itemize}
 \item $\mathrm{5^2S_{1/2}}\rightarrow \mathrm{5^2P_{3/2}}$ : 780 nm ($\mathrm{D}_2$ line)
 \item $\mathrm{5^2S_{1/2}}\rightarrow \mathrm{5^2P_{1/2}}$ : 795 nm ($\mathrm{D}_1$ line)
 \item $\mathrm{5^2P_{3/2}}\rightarrow \mathrm{5^2D_{5/2}}$ : 776 nm
 \item $\mathrm{5^2P_{3/2}}\rightarrow \mathrm{5^2D_{3/2}}$ : 776 nm
 \item $\mathrm{5^2P_{1/2}}\rightarrow \mathrm{5^2D_{3/2}}$ : 762 nm
 \item $\mathrm{5^2P_{3/2}}\rightarrow \mathrm{7^2S_{1/2}}$ : 741 nm
\end{itemize}

With the atomic state evolution calculated for a given field $\mathcal{E}(r,z,t)$, the corresponding atomic polarization envelope required for Eq. \ref{waveeq} is computed as the expectation value of the dipole operator: 
 \begin{equation}
\tilde{\mathcal{P}}= \mathfrak{F}\left\{\mathcal{N}\sum_{kl}\alpha_k^*\alpha_l d_{kl}\right\}.
\label{eq_P}
\end{equation}

The third term in the RHS of Eq. \ref{waveeq}, accounts for energy loss due to ionization. It ensures that the energy removed from the field matches the total photon energy required to ionize the atom:
\begin{equation}
\tilde{\mathcal{Q}}=\mathfrak{F}\left\{\sum_j n_j 
\frac{\Gamma_j |a_j|^2}{\mathcal{E}^*}\right\}.
\label{eq_Q}
\end{equation}
where $n_j$ is the number of photons required to ionize from state $|j\rangle$. Note that here $n_j$ is not an integer, but an expectation value that is determined during the calculation of the ionization rates $\Gamma_j$ and includes the possibility of ionization with more than the minimum number of photons (above threshold ionization) if the light intensity is high enough. However, in our simulations they are always practically equal to the minimum number.  

The final term, $\tilde{\mathcal{R}}$, describes the contribution of the free-electron plasma to dispersion. It is proportional to the ionization probability:
\begin{equation}
\tilde{\mathcal{R}}=\mathfrak{F}\left\{
(1-\sum_j |a_j|^2)\mathcal{E}
\right\}.
\label{eq_R}
\end{equation}

\subsection{Numerical procedure}

Equation \ref{waveeq}  been solved using a split operator technique. The diffraction part (first term on the RHS) was advanced along the $z$ axis using a Crank-Nicolson implicit scheme. At the outer boundary $r=r_{max}$ of the numerical integration, a {\em perfectly matched layer} was implemented to prevent artificial reflections from the boundary during the long propagation calculations. The material parts (terms 2-4 of the RHS) were advanced along $z$ using a predictor-corrector stepper. Equations 
\ref{schrodinger} have been solved in the time domain using a fifth-order, adaptive Dormand-Prince algorithm at each spatial point. With the atomic probability amplitudes calculated, the material parts were computed according to Eqs. \ref{eq_P}, \ref{eq_Q} and \ref{eq_R}. 

The pulses entering the rubidium vapor at $z=0$ were taken to be a pure Gaussian beam defined with the waist parameter $w_0$ and waist location $z_0$. The initial envelope time dependence was taken to be $\mathrm{sech}(t/\tau)$ with $\tau=68.08\mathrm{~fs}$ that yields a 120 fs duration pulse.


\section{Simulation results}

\subsection{Pulse input energy scan}
Equations \ref{waveeq} and \ref{schrodinger} were solved using a cloud-based high-performance computing cluster to investigate the propagation dynamics of high-power laser pulses with varying central wavelengths. First, simulations were conducted for three different wavelengths—$\lambda = 750$ nm, $\lambda = 780$ nm, and $\lambda = 810$ nm—across a range of initial pulse energies $E_{\text{in}}$. The remaining simulation parameters were consistent with those used in some previous experiments and numerical studies \cite{Demeter2021}, namely: vapor density $\mathcal{N} = 6.6 \times 10^{14}\mathrm{cm}^{-3}$, beam waist parameters $w_0 = 1.506\mathrm{~mm}$ and $z_0 = 7.92~\mathrm{m}$. Pulse duration was $T = 120~\mathrm{fs}$, which yields a transform-limited bandwidth of $\sim$ 7.5 nm.

\begin{figure}[htb]
\includegraphics[width=1\columnwidth]{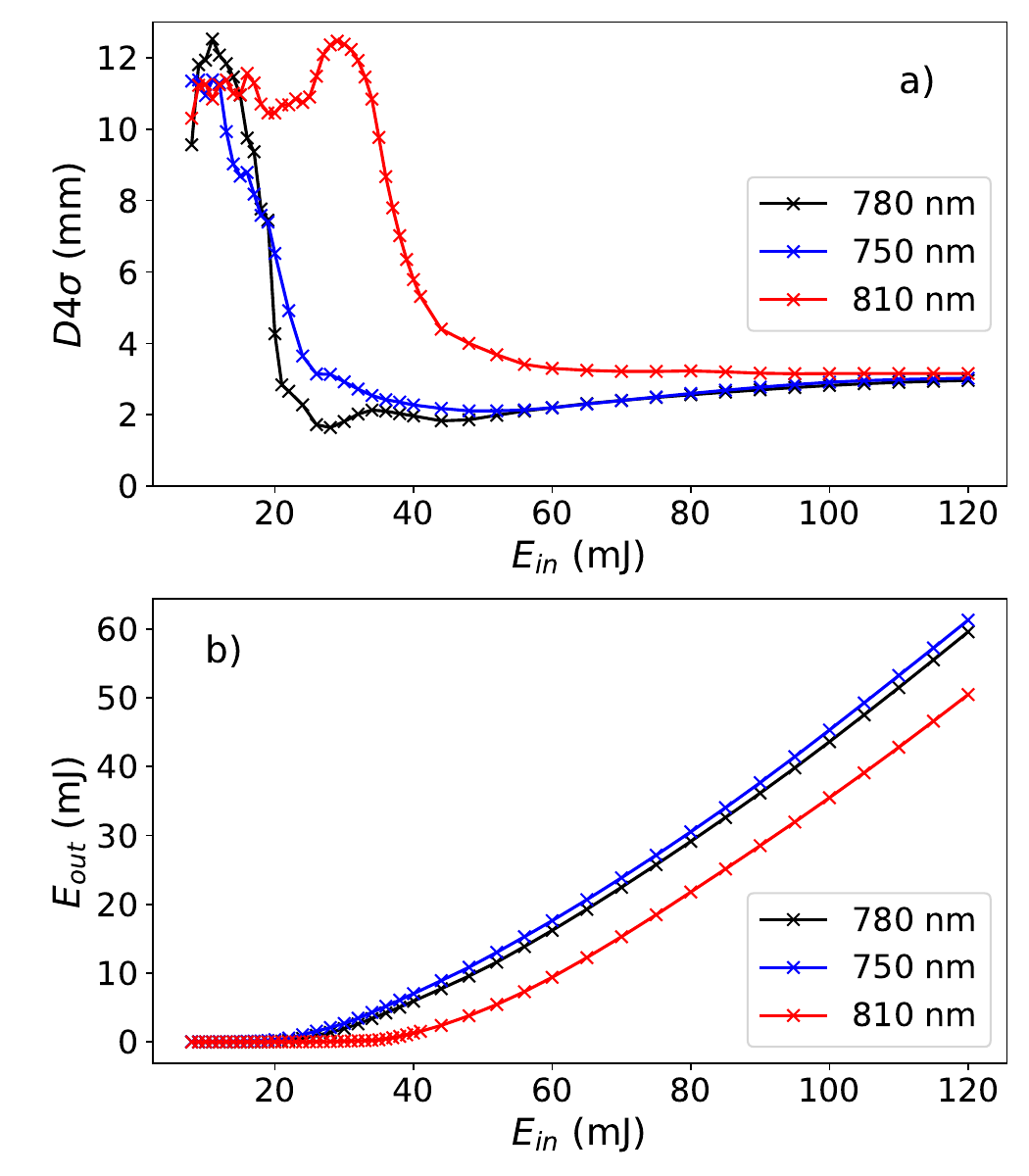}
\caption{Transmitted pulse properties at $z=10$ m propagation distance as a function of input pulse energy $E_{in}$ for the different central wavelengths. a) Transmitted pulse $D4\sigma$ width. b) Transmitted energy $E_{out}$.  }
\label{fig_transmittedpulses}       
\end{figure}

To compare pulse propagation at the resonant wavelength (780 nm) with that at detuned values of ±30 nm, we analyzed two experimentally measurable quantities: the {\em transmitted pulse width}, defined as the $D4\sigma$ width of the pulse at a propagation distance of $z = 10~\mathrm{m}$, and the {\em transmitted energy} $E_{\text{out}}$, which quantifies the remaining energy in the pulse at this location. The evolution of these parameters as a function of $E_{\text{in}}$ was previously validated against experiments for 780 nm and 810 nm pulses \cite{Demeter2024}, with the simulation capturing the qualitative trends well. Figures \ref{fig_transmittedpulses}a and \ref{fig_transmittedpulses}b display these dependencies.

In the low-energy regime (the so-called sub-threshold domain), the transmitted pulses are wide and exhibit very low transmitted energy. This corresponds to an almost complete absorption of the pulse by the vapor before $z=10$ m propagation distance is reached and the termination of the plasma channel well before that. As $E_{\text{in}}$ increases, there is an abrupt narrowing of the transmitted pulse, accompanied by a slow increase in $E_{\text{out}}$. Notably, the behavior of 750 nm pulses closely resembles that of the resonant 780 nm pulses: the onset of energy transmission occurs at lower input energies compared to 810 nm pulses, and the transmitted energy is also higher for a given $E_{\text{in}}$. The transmitted pulse width for 750 nm is similarly close to that of the 780 nm case, exhibiting a minimum width above the breakthrough point before increasing again (this has been called the confined beam domain for 780 nm pulses earlier \cite{Demeter2021} ). These results indicate that 750 nm pulses undergo a propagation regime transition at similar energy thresholds and follow comparable confinement dynamics as the 780 nm wavelength pulses.

Additionally, the spatial characteristics of the plasma channel were analyzed using definitions inspired by prior schlieren imaging measurements reported in \cite{Demeter2024}. Two parameters were introduced to quantify the plasma structure: the plasma channel radius $r_p(z)$ and the plasma sheath width $w_p(z)$. These are defined as follows:
\begin{equation}
 r_p(z) = \int_0^\infty P_{ion}(r,z)dr
\end{equation}
\begin{equation}
 w_p(z) = \int_0^\infty |P_{ion}(r,z)-H(r_p-r)|dr
\end{equation}
where $P_{\text{ion}}(r,z)$ is the local ionization fraction (i.e., the probability that an atom is ionized after passage of the laser pulse), and $H(x)$ is the Heaviside step function. The quantity $r_p(z)$ represents the equivalent step-function radius, and $w_p(z)$ quantifies the width of the transition zone between the fully ionized plasma core ($P_{\text{ion}} = 1$) and the surrounding neutral vapor ($P_{\text{ion}} = 0$). These definitions offer increased robustness over fitting-based methods  applied to numerically obtained ionization profiles.

\begin{figure}[htb]
\includegraphics[width=1\columnwidth]{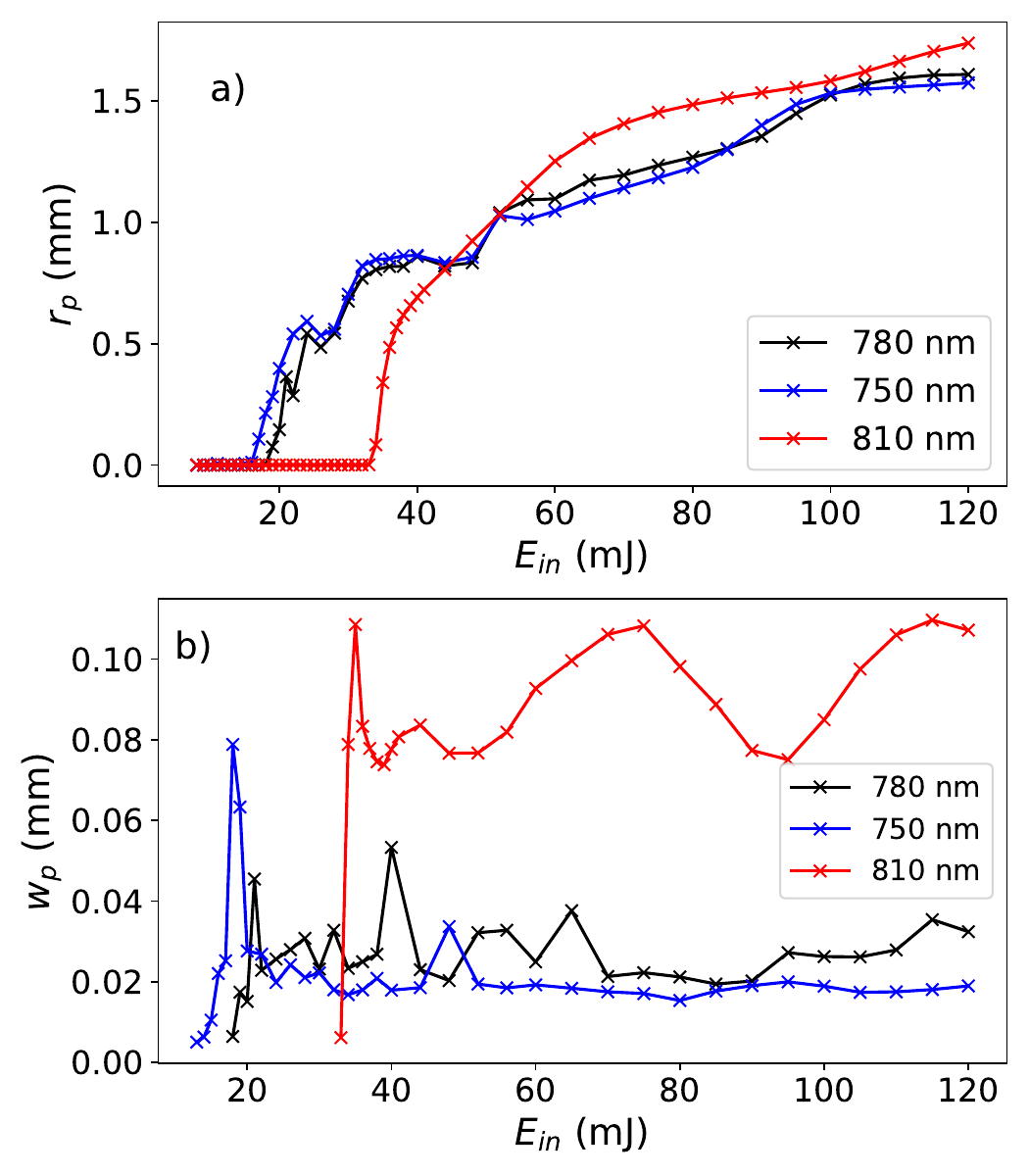}
\caption{a) Plasma channel radius $r_p$ and b) plasma channel sheath width $w_p$ at a propagation distance $z=10$ m as a function of input pulse energy $E_{in}$ for the different central wavelengths.  }
\label{fig_plasmaproperties}       
\end{figure}

Figures \ref{fig_plasmaproperties}a and \ref{fig_plasmaproperties}b present the calculated values of $r_p$ and $w_p$ at $z = 10~\mathrm{m}$. Consistent with earlier findings, the plasma radius $r_p$ begins to increase at lower input energies for 750 nm and 780 nm pulses compared to 810 nm. Furthermore, the plasma sheath width $w_p$ is significantly smaller—and nearly identical—for the 750 nm and 780 nm cases, indicating more confined plasma channel formation relative to 810 nm pulses.
Thus these results suggest that sub-resonant wavelength pulses at 750 nm are focused and guided by the vapor medium about as effectively as resonant 780 nm pulses, and significantly more so than the 810 nm case.

\begin{figure}[htb]
\includegraphics[width=1\columnwidth]{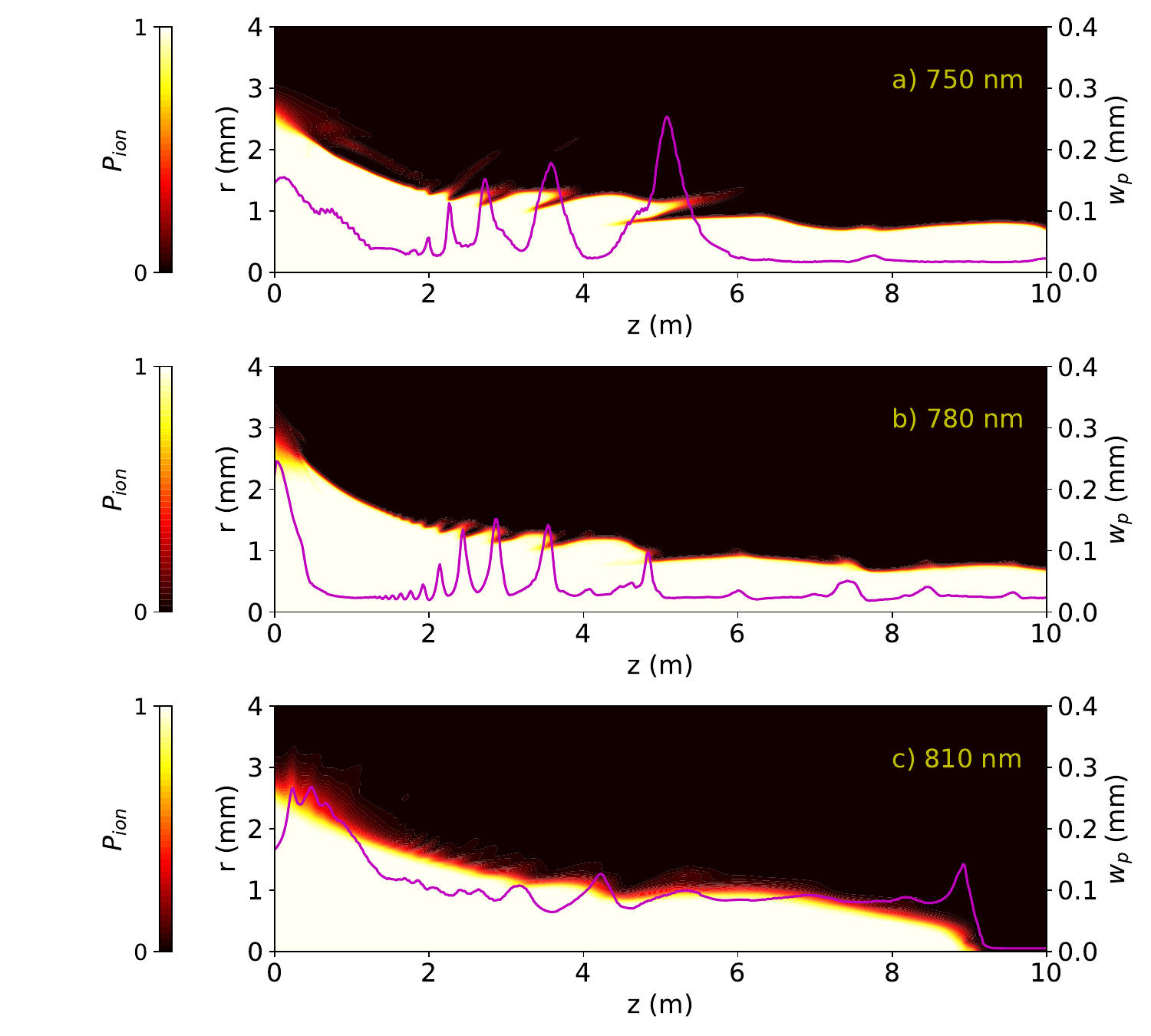}
\caption{Contour plots of the ionization probability $P_{ion}(r,z)$ showing the plasma channels created by the laser pulses in the vapor (left $y$-axis). Line plots of the plasma sheath width $w_p$ along the vapor (magenta line, right $y$-axis). $E_{in}=30$ mJ. }
\label{fig_plasmachannels1}       
\end{figure}

\begin{figure}[htb]
\includegraphics[width=1\columnwidth]{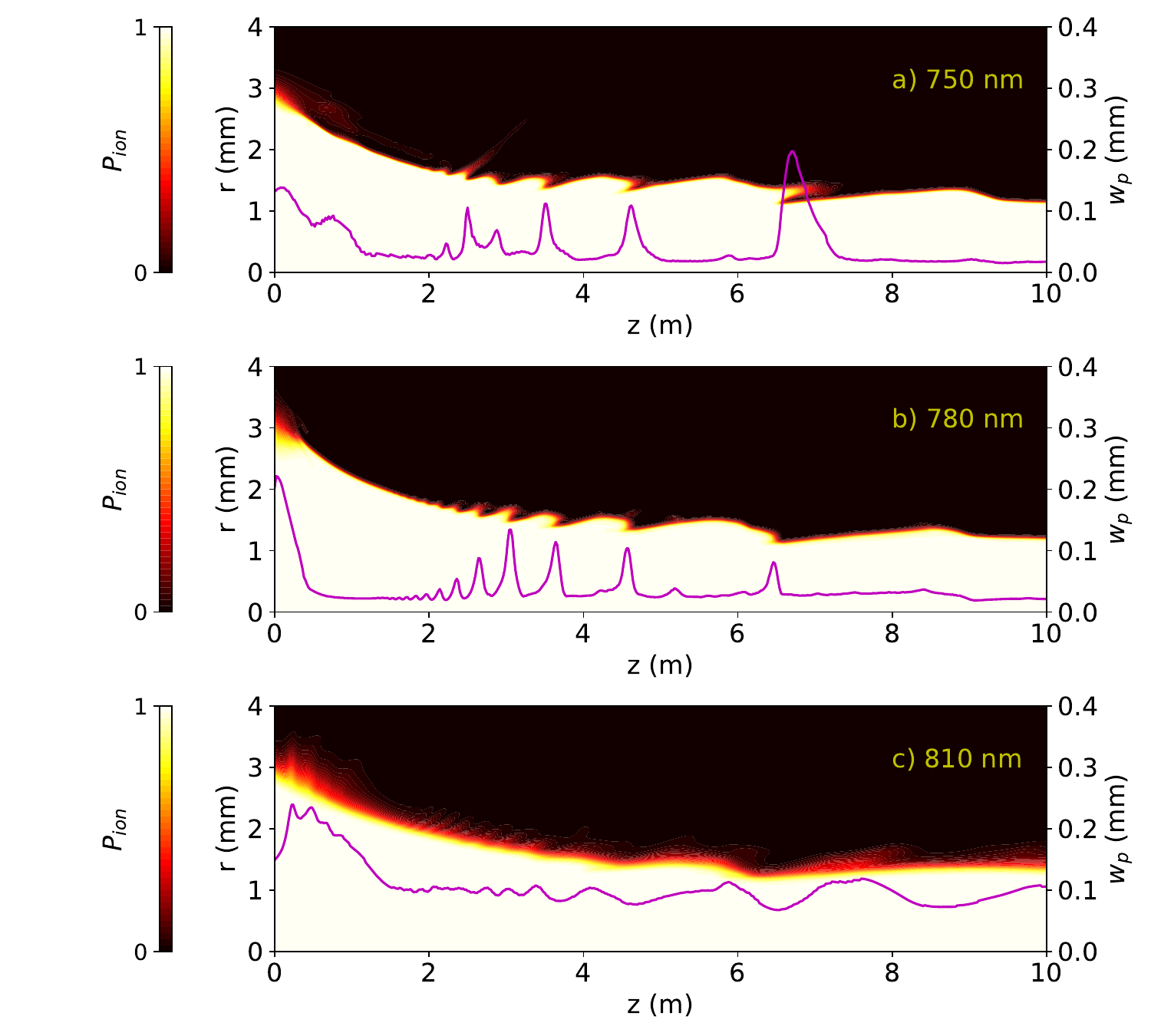}
\caption{Contour plots of the ionization probability $P_{ion}(r,z)$ showing the plasma channels created by the laser pulses in the vapor (left $y$-axis). Line plots of the plasma sheath width $w_p$ along the vapor (magenta line, right $y$-axis). $E_{in}=70$ mJ.  }
\label{fig_plasmachannels2}       
\end{figure}

Further insights into the propagation dynamics can be gained by examining the evolution of the plasma channel along the propagation axis.
Figures \ref{fig_plasmachannels1} and \ref{fig_plasmachannels2} present the simulated ionization fraction $P_{\text{ion}}(r,z)$—i.e., the plasma density normalized to the background vapor density $\mathcal{N}$—as a function of radial and axial position, for the three central wavelengths and two representative input pulse energies. The plasma density is shown as a contour plot, while the plasma sheath width $w_p(z)$ is overlaid as a magenta curve (referencing the right vertical axis).

The contour plots of $P_{\text{ion}}(r,z)$ reveal that after an initial focusing phase over the first $\sim$2 meters, the plasma channels formed by the 750 nm and 780 nm pulses exhibit similar periodic dynamics. Regions where the plasma channel becomes narrower and the laser field more focused are followed by an expansion of the laser beam and the corresponding slight widening of the plasma channel. These modulations result from repeated self-focusing of the ionizing pulse and are a hallmark of filamentary propagation. Overall, the plasma channel becomes narrower and narrower as the pulse loses energy during propagation.
In contrast, such oscillatory behavior is significantly less pronounced for the 810 nm pulses.

Moreover, the sheath width $w_p(z)$ tends to be smaller for the shorter wavelengths (750 nm and 780 nm), indicating a sharper ionization boundary. However, $w_p(z)$ displays more pronounced fluctuations at these wavelengths around the turning points where the plasma channel transitions from contraction to expansion. These spikes are broader and reach higher amplitudes for the 750 nm case, suggesting that the transition between focusing and defocusing is most abrupt for the resonant, 780 nm pulses. 
This behavior indicates that while 750 nm pulses closely replicate the general plasma channel structure of resonant pulses—with similar $r_p$ and $w_p$ values at $z=10~\mathrm{m}$ (see Fig. \ref{fig_plasmaproperties})—the nonlinear self-focusing dynamics are somewhat less effective.

In contrast, 810 nm pulses generate smoother, less modulated plasma channels with broader and more stable sheath regions. In addition, for $E_{\text{in}} = 30~\mathrm{mJ}$ (Fig. \ref{fig_plasmachannels1}c), the plasma channel formed by the 810 nm pulse does not extend over the full 10-meter propagation distance. Energy is fully depleted and the pulse ``crashes''.

\begin{figure}[htb]
\includegraphics[width=1\columnwidth]{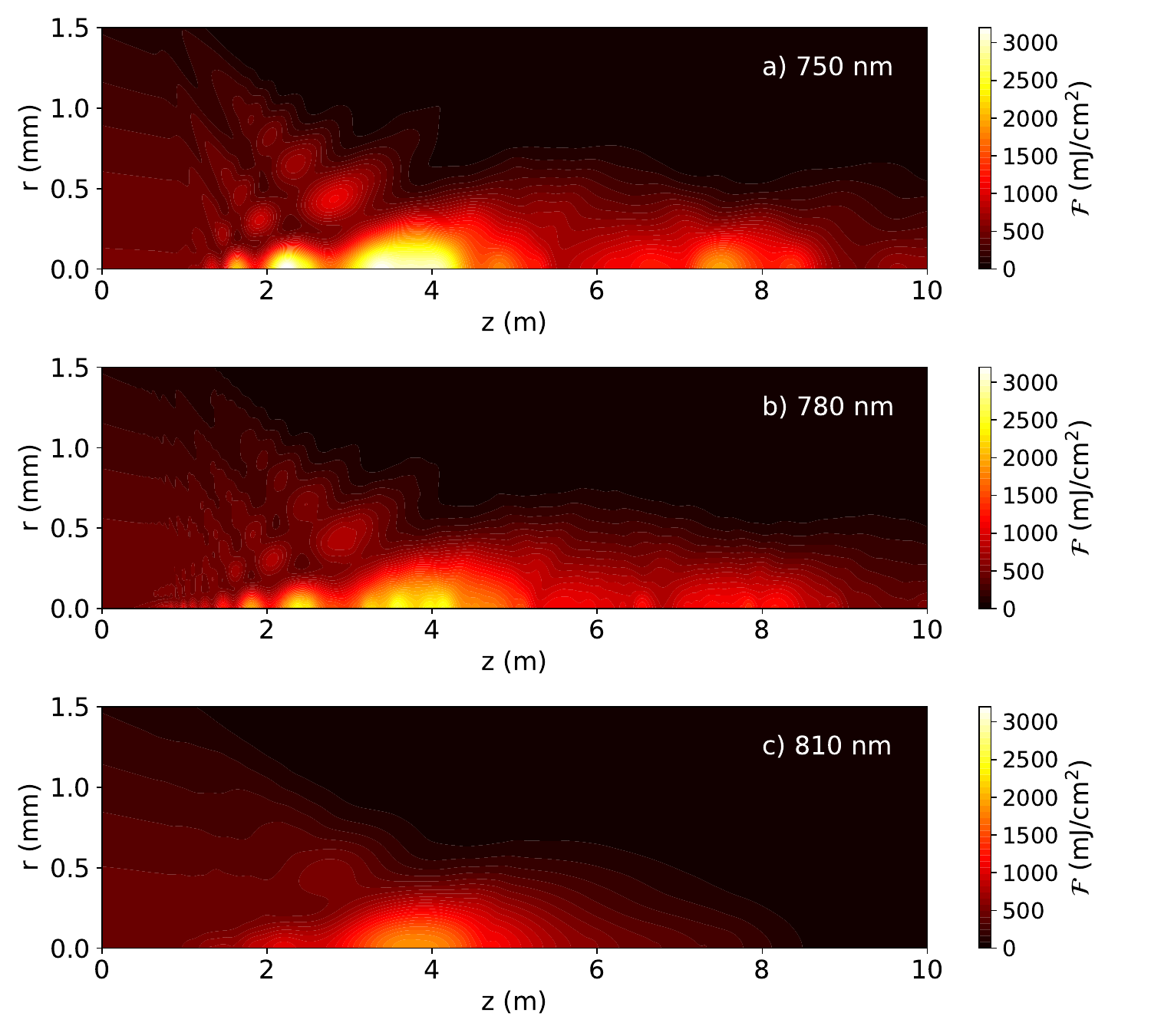}
\caption{ Contour plots of the radiant fluence $\mathcal{F}(r,z)$ of the laser pulses in space, for $E_{in}=30$ mJ. Color scale of the three plots is the same.}
\label{fig_fluences}       
\end{figure}

{ Additionally, Figure \ref{fig_fluences} shows the radiant fluence $\mathcal{F}(r,z)$ in space for the three wavelengths. From these figures it is visible that the 750 nm and 780 nm pulses both undergo a series of focusings where radiant fluence increases sharply, but the 810 nm pulse has only a single maximum of the fluence along the propagation axis. The behavior of the fluence for the two lower wavelengths is quite similar both in structure and amplitude, but in the vicinity of the maximum fluence around $z=3$ - $4$ m, the 780 nm pulse possesses an additional finer structure of local maxima. Overall these figures support fully the conclusions drawn from the plasma channel profiles on Figs. \ref{fig_plasmachannels1} and \ref{fig_plasmachannels2}.}

\subsection{Pulse wavelength scan}

To obtain a more comprehensive understanding of wavelength-dependent propagation behavior, we conducted an additional set of simulations scanning the central wavelength $\lambda$ of the laser pulses in the vicinity of the $\mathrm{D}_2$ resonance. For this study, we fixed the input pulse energy at $E_{\text{in}} = 3.5~\mathrm{mJ}$ and used the following simulation parameters: vapor density $\mathcal{N} = 2 \times 10^{14}\mathrm{~cm}^{-3}$, beam waist $w_0 = 1.274\mathrm{~mm}$, and focal position $z_0 = 0~\mathrm{m}$. { These physical parameters were chosen so that the ``crash'' of the laser pulse, the position where energy loss prevents further ionization and the plasma channel ends, can be observed within the integration domain for all wavelengths. }

\begin{figure}[h!tb]
\centerline{\includegraphics[width=0.8\columnwidth]{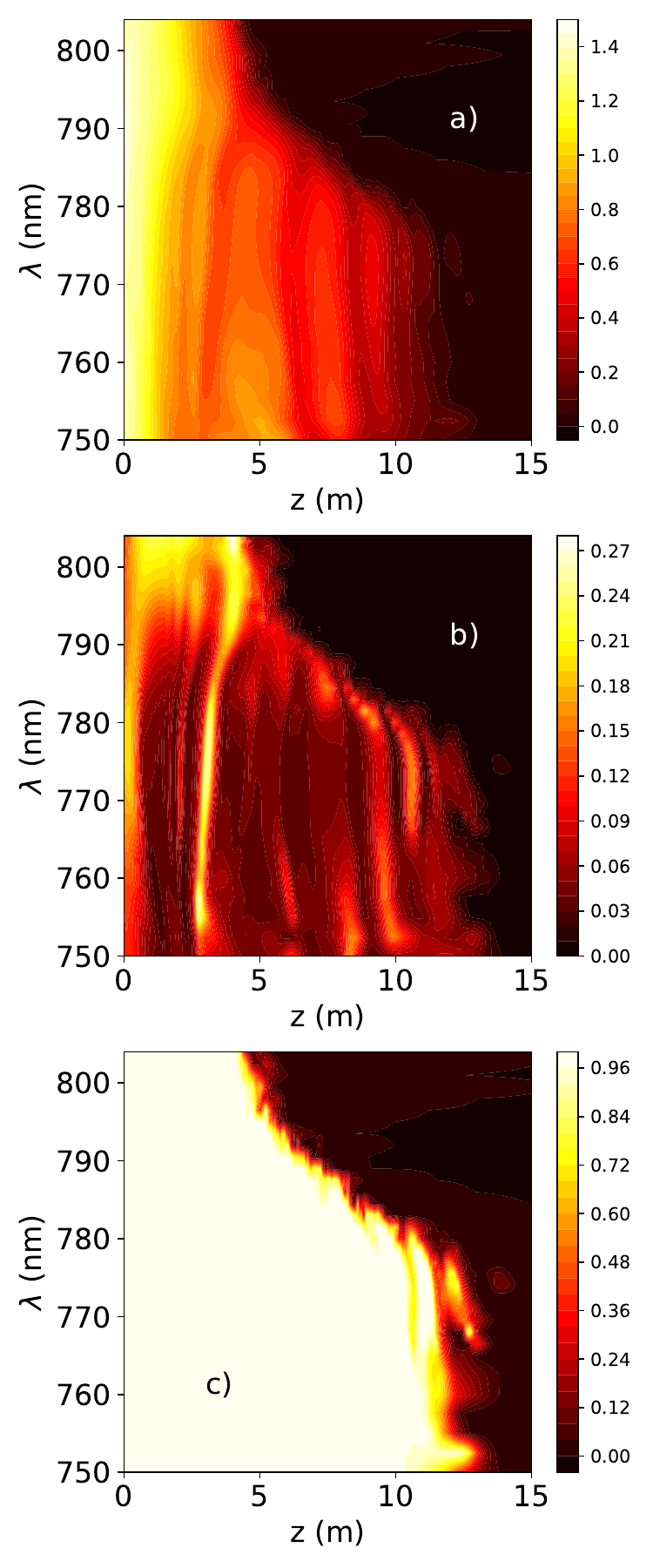} }
\caption{Contour plots of a) the plasma radius $r_p$ in mm, b) plasma sheath width $w_p$ in mm and c) on-axis ionization fraction $P_{ion}(r=0,z)$ as a function of propagation distance $z$ and pulse central wavelength $\lambda$.   }
\label{fig_lambdascan}       
\end{figure}

Figure \ref{fig_lambdascan} a) presents the plasma channel radius $r_p(z)$ as a function of both propagation distance $z$ and central wavelength $\lambda$. The results reveal a pronounced asymmetry in behavior around the $\mathrm{D}_2$ resonance at 780 nm. For sub-resonant wavelengths ($\lambda < 780~\mathrm{nm}$), the plasma channel radius fluctuates along $z$ but remains relatively insensitive to changes in $\lambda$. In contrast, for super-resonant wavelengths ($\lambda > 780~\mathrm{nm}$), the propagation distance over which a significant plasma channel is sustained decreases rapidly as $\lambda$ increases.

This asymmetry is further illustrated in Figs. \ref{fig_lambdascan} b), and \ref{fig_lambdascan} c). Fig. \ref{fig_lambdascan} b) shows the plasma sheath width $w_p$\texttwelveudash   the narrow, highly fluctuating behavior at and below 780 nm slowly transforms to the wider, more uniform behavior as the 810 nm wavelength is approached. Fig. \ref{fig_lambdascan} c) shows the on-axis ionization fraction $P_{\text{ion}}(r = 0, z)$. The length of the fully ionized region decreases significantly with increasing $\lambda$ above 780 nm, while remaining comparatively stable for wavelengths below the resonance. These results are somewhat surprising, given that the $\mathrm{D}_1$ transition to the $\mathrm{5^2P_{1/2}}$ state at 795 nm is the next strongest dipole-allowed transition from the ground state of rubidium after the $\mathrm{D}_2$ line. Nevertheless, the simulation suggests that the nonlinear response is dominated by the strong near-resonant interaction at 780 nm and drops off asymmetrically for longer wavelengths, despite the presence of the $\mathrm{D}_1$ line.

\section{Discussion}

The observed asymmetry in the propagation around the $\mathrm{D}_2$ resonance is not completely unexpected, there are multiple factors that may play a role:
\begin{itemize}
 \item Multiphoton ionization (MPI) rates depend on the $\lambda$ wavelength: $W\sim I^N\lambda^{2N}$ (here $I$ is the light intensity, $N$ the number of photons required for ionization). This means that for the 5S ground state, the rate is about 20\% larger for 810 nm light, than for 780 nm light. Similarly, for 750 nm light it is about 20\% smaller. Thus for lower wavelengths, atomic polarization due to the valence electron will persist for longer time in the case of shorter wavelengths.

 \item The valence electron of Rb has several excited states with dipole allowed transitions between them that lie within, or close to the 780-750 nm range. These are the $\mathrm{5^2P_{3/2}} \rightarrow \mathrm{5^2D_{3/2}}$ and $\mathrm{5^2P_{3/2}} \rightarrow \mathrm{5^2D_{5/2}}$ transitions near 776 nm, the $\mathrm{5^2P_{1/2}} \rightarrow \mathrm{5^2D_{3/2}}$ transition at 761 nm, and the $\mathrm{5^2P_{3/2}}\rightarrow \mathrm{7^2S_{1/2}}$ one at 741 nm. These transitions can be strongly driven in the plasma-vapor boundary layer where excited-state populations are significant, as possibly the the two-photon $\mathrm{5^2S_{1/2}} \rightarrow \mathrm{7^2S_{1/2}}$ transition at ~760 nm wavelength. These transitions
 can have a substantial effect on the local refractive index during propagation. Since the transform limited bandwidth of the 120 fs duration pulse is approximately $\Delta \lambda \approx 7.5~\mathrm{nm}$, these transitions may contribute to atomic polarization simultaneously during the interaction with powerful light fields.

 \item Even though traditional linear susceptibility is not a valid description of the optical response at the timescales and intensities discussed here, it is instructive to recall that around an atomic resonance, the anomalous dispersion curve predicts a negative refractive index change $\Delta n < 0$ for blue-detuned wavelengths $\lambda < \lambda_0$ and a positive one $\Delta n > 0$ for red-detuned wavelengths $\lambda > \lambda_0$. This means, that in the plasma sheath layer, where the density of (still un-ionized) atoms increases radially outward from the core, the radial dependence of the refractive index works as a focusing medium for blue-detuned $\lambda < \lambda_0$ wavelengths and as a de-focusing one for red-detuned $\lambda > \lambda_0$ wavelengths. 

\end{itemize}

Accurately quantifying the contribution of each of the above factors is challenging, as they depend sensitively on the local pulse intensity and on the detailed temporal and spatial structure of the pulse. These evolve during propagation through self-focusing and plasma formation. Nevertheless, it is instructive to calculate the atomic polarization for a given specific laser pulse envelope while varying the wavelength. This has been done by solving Eqs. \ref{schrodinger} and computing the expectation value of the dipole operator $\mathcal{P}_a=\sum_{kl}\alpha_k^*\alpha_l d_{kl}$.

\begin{figure}[h!tb]
\centerline{\includegraphics[width=0.8\columnwidth]{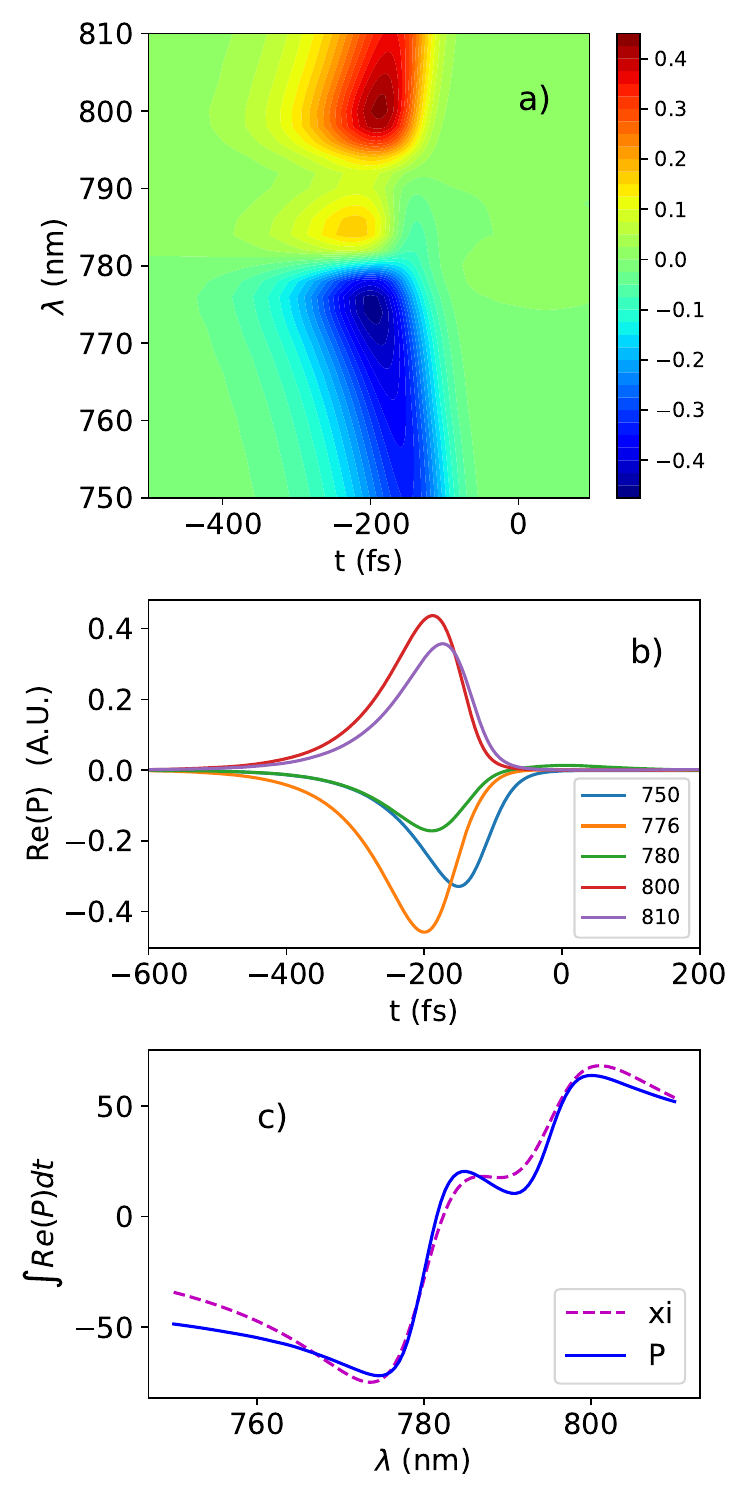} }
\caption{Plots of the real part of the atomic polarization $\Re(\mathcal{P}_a)$ measured in atomic units $e\cdot a_0$ during a 120 fs duration sech laser pulse with a peak intensity of $7\cdot 10^{10} \mathrm{~W/cm^2}$.  a) Contour plot of $\Re(\mathcal{P}_a)$ as a function of time and pulse wavelength. b) Line plots of $\Re(\mathcal{P}_a)$ for selected wavelengths. c) The time integral  $\int\Re(\mathcal{P}_a(t))dt$ as a function of pulse wavelength (solid blue line) and a best fit line of two Lorentzian lineshape functions (dashed magenta line).}
\label{fig_polarization}       
\end{figure}

Figure \ref{fig_polarization} shows some plots of the real (in-phase) part of the polarization 
$\Re(\mathcal{P}_a)$ which is responsible for the refractive index change due to atomic transitions. 
Fig. \ref{fig_polarization} a) is a contour plot of the wavelength and time dependence of $\Re(\mathcal{P}_a)$, while Fig. \ref{fig_polarization} b) shows line plots of the same quantity for selected wavelengths.
One can see from these two figures, that:
\begin{itemize}
 \item $\Re(\mathcal{P}_a)$ is negative for $\lambda\leq780$ nm, which serves to reduce the refractive index. The maximum of the magnitude of $\Re(\mathcal{P}_a)$ in this region is around 776 nm, hinting that this is where focusing by the medium in the sheath layer should be the strongest.
 
 \item $\Re(\mathcal{P}_a)$ becomes positive just above the $\mathrm{D}_2$ resonance, which serves to increase the refractive index. The maximum of $\Re(\mathcal{P}_a)$ is attained at around $\lambda=800$ nm, just above the 795 nm $\mathrm{D}_1$ resonance, hinting that this is where defocusing by the medium in the sheath layer should be the strongest.
 
 \item There is a slight modulation of $\Re(\mathcal{P}_a)$ with $\lambda$ between the stronger $\mathrm{D}_2$-transition and the weaker $\mathrm{D}_1$-transition wavelengths.
 
 \item The polarization becomes zero for all wavelengths due to ionization well before $t=0$, the center of the pulse envelope. Wavelengths close to the $\mathrm{D}_2$ and $\mathrm{D}_1$ resonances are ionized faster as the process is expedited by resonant intermediate states. 

\end{itemize}

The final plot, Fig. \ref{fig_polarization} c) depicts the time integral of $\int\Re(\mathcal{P}_a)dt$ with a solid blue line, which is the overall in-phase response of the atom to the driving field in this transient domain. The wavelength-dependence of this curve is strikingly similar to the superposition of two anomalous dispersion curves around the $\mathrm{D}_2$ and $\mathrm{D}_1$ resonances as given by the classical Lorentz oscillator model \cite{SalehTeich}. Indeed, a numerical fit to the data of the form: 
\begin{eqnarray}
 \chi & = & A_1\chi_1+A_2\chi_2 \mathrm{~~~with}\\
 \chi_i & = & A_i \frac{f-f_{0i}}{(f-f_{0i})^2 + (\Delta f/2)^2}
\end{eqnarray}
where $f = c/\lambda$ and $f_{01}=c/780 \mathrm{~nm}$, $f_{02}=c/795 \mathrm{~nm}$, is also plotted on Fig. \ref{fig_polarization} c) with a dashed line. The similarity between the curves suggests that even though we are considering a short-timescale, transient atomic response in the presence of powerful fields and ionization, the two single-photon resonances from the ground state still have a major influence on the optical response. Clearly, wavelength-dependent ionization rates and transitions between excited states will also have an effect, so the correspondence between the two curves is not perfect.
While our calculation here has been done with a pure sech pulse shape and a relatively low amplitude, the results probably apply to some measure anywhere in the plasma sheath, whatever the pulse shape that developed during propagation. 

 { For the most part, pulses are intense enough for multiphoton ionization to take place, so most of the energy is lost during propagation from the field via this channel for all wavelengths. The differences arise predominantly near the low intensity edge of the pulse (where the plasma sheath is formed) and near the region where the propagation ends due to energy depletion (where the pulse ``crashes''). In these low-intensity regions, resonant single-photon absorption is expected to be dominant on the $\mathrm{D}_2$ and $\mathrm{D}_1$ lines for wavelengths around and above 780 nm. While the closer line would clearly dominate in this wavelength-region, due to the high bandwidth of the pulses none of these wavelengths are completely ``non-resonant'' on the red side of the $\mathrm{D}_2$ line, even the 810 nm wavelength will experience considerable absorption on the $\mathrm{D}_1$ line. Wavelengths on the blue side on the other hand will be strongly focused by anomalous dispersion-like behavior, so multiphoton absorption and ionization will persist much longer. Thus, all wavelengths considered will eventually be absorbed completely mostly due to the interaction with the two single-photon transitions from the ground state.}

Finally, we also note that in the wavelength range investigated, the optical nonlinearities of the vapor due to the Rb I ion core are negligible, owing to the very low vapor density\textendash several orders of magnitude below standard atmospheric density. As a result, once ionization occurs, the medium becomes effectively transparent to the laser field. This enables a very efficient guiding of the laser energy inside the plasma channel core when the optical response of the atoms in the sheath layer results in focusing.

\section{Summary}

We investigated how the propagation of TW power laser pulses in rubidium vapor depends on 
wavelength, focusing on wavelengths near the 780 nm $\mathrm{D}_2$ and 795 nm $\mathrm{D}_1$
resonance lines. To model this interaction, we derived a set of equations that describe the 
transient optical response of the atoms both at exact resonance and at nearby off-resonant 
wavelengths. Notably, our model incorporates several transitions of the rubidium valence 
electron between excited states with resonant wavelengths close to the $\mathrm{D}_2$ and $\mathrm{D}_1$ lines.

Using cloud-based computational resources, we solved these equations for three distinct wavelengths. We analyzed how the final pulse energy and the transverse beam width at a propagation distance of $z=10$ meters depend on the input energy. Additionally, we calculated the plasma channel radius and the width of the surrounding plasma boundary layer at this distance. Further simulations were conducted with fixed pulse energy while varying the pulse wavelength. We examined how the plasma channel characteristics\textendash specifically, the channel radius and boundary layer thickness\textendash change with both wavelength and propagation distance.

Our findings reveal a marked asymmetry in propagation behavior on either side of the 780 nm $\mathrm{D}_2$ resonance. For wavelengths shorter than 780 nm, the propagation dynamics closely resemble those at exact resonance. The transmitted pulse energy and beam width show nearly identical trends, and the plasma channel exhibits a sharply defined boundary. The primary difference is that the periodic switching between self-focusing and beam expansion occurs more rapidly in the exactly resonant case.
In contrast, for wavelengths longer than 780 nm, the propagation gradually shifts toward behavior observed at the far-off-resonant 810 nm case. This regime is characterized by weaker self-focusing effects, a more uniform and significantly broader plasma sheath, and lower transmitted pulse energy. The transition between these regimes is fairly smooth. 

We have calculated the transient optical response of the rubidium atom to an ionizing, ultrashort laser pulse. The wavelength dependence of the real part of the atomic polarization was compared with that predicted by the superposition of classical Lorentzian line shapes for the two principal resonances. This comparison aids in explaining the pronounced difference in plasma sheath width observed for wavelengths above and below the 780 nm $\mathrm{D}_2$ transition. Our findings indicate that, within the investigated wavelength range, pulse propagation is governed primarily by the two dominant resonance lines of rubidium, with additional influence from excited-state transitions and wavelength-dependent ionization rates. In particular, transitions near 776 nm, 762 nm, and 740 nm appear to contribute to the observed behavior.


\section*{Acknowledgment}
The research was supported by the Hungarian National Research, Development and Innovation Office (NKFIH) under the contract number 2021-4.1.2-NEMZ\_KI-2024-00041. 
On behalf of Project Awakelaser we are grateful for the usage of HUN-REN Cloud \cite{Heder2022} which helped us achieve the results published in this paper.

\clearpage

\bibliography{/home/gdemeter/fiz/manuscript/bibliography_pulseprop}

\bibliographystyle{apsrev4-1}


\end{document}